\documentstyle[12pt,epsfig]{article}
\begin{document}
{\large \sf

{\sf
\title{
{\normalsize
\begin{flushright}
%CU-TP
\end{flushright}}
Convergent Iterative Solutions of Schroedinger Equation for a
Generalized Double Well Potential\thanks{Work supported in part by
the U.S. Department of Energy}}

\author{
R. Friedberg$^{1}$, T. D. Lee$^{1,2}$ and W. Q. Zhao$^{2,3}$\\
{\small \it 1. Physics Department, Columbia University, New York, NY 10027, USA}\\
{\small \it 2. China Center of Advanced Science and Technology (CCAST)}\\
{\small \it         (World Lab.), P.O. Box 8730, Beijing 100080,  China}\\
{\small \it 3. Institute of High Energy Physics, Chinese Academy of
Sciences, Beijing 100039, China}  } \maketitle
%\date{}

\begin{abstract}
We present an explicit convergent iterative solution for the
lowest energy state of the Schroedinger equation with a
generalized double well potential
$V=\frac{g^2}{2}(x^2-1)^2(x^2+a)$. The condition for the
convergence of the iteration procedure and the dependence of the
shape of the groundstate wave function on the parameter $a$ are
discussed.
\end{abstract}

\vspace{1cm}

{\normalsize ~~~~PACS{:~~11.10.Ef,~~03.65.Ge}}

\newpage

\section*{\bf 1. Introduction}
\setcounter{section}{1}
\setcounter{equation}{0}

This paper is stimulated by an interesting question raised by Roman
Jackiw[1] concerning the extent of validity of the convergent
iterative method[2-5] that we have developed for the $N$-dimensional
generalization of the double well potential
$$
V=\frac{g^2}{2}(r^2-r_0^2)^2.\eqno(1.1)
$$
More specifically, whether the method is equally applicable to a
different sombrero-shaped potential
$$
V=\frac{g^2}{2}(r^2-r_0^2)^2(r^2+2r_0^2).\eqno(1.2)
$$
The latter has several new features[1,6]. When $g^2=1$ and
$$
r_0^4=\frac{1}{3}(2+N),\eqno(1.3)
$$
the groundstate wave function $\psi$ is simply
$$
e^{-r^4/4}\eqno(1.4)
$$
which peaks at $r=0$. Yet for $g^2>>1$, the maximum of $\psi$ has to
be near $r=r_0$. In particular, for a one-dimensional problem, as
$g^2$ increases from $1$, there would be a critical point when
$\psi$ changes from having a single maximum at the origin to one
with double peaks. Thus, it is of interest to examine whether our
convergent iterative method works for $g^2=1$ as well as for $g^2$
larger than $1$. The purpose of this paper is to show that this is
indeed the case.

To simplify our discussions, we examine only the one-dimensional
case in this paper. Let $\psi(x)$ be the groundstate wave function
of the Schroedinger equation

$$
\bigg[-\frac{1}{2}\frac{d^2}{dx^2}+V(x)\bigg]\psi(x)=E\psi(x)\eqno(1.5)
$$
with
$$
V(x)=\frac{g^2}{2}(x^2-1)^2(x^2+a)\eqno(1.6)
$$
and
$$
a>0.\eqno(1.7)
$$
We note that in one dimension, the potential (1.2)-(1.3) is a
special case of (1.6) with $a=2$. For convenience of nomenclature,
in dimension $>1$ we call (1.1) the sombrero potential, and (1.2)
the generalized sombrero potential; in one dimension, we call
$$
\frac{g^2}{2}(x^2-1)^2\eqno(1.8)
$$
the double well potential and (1.6) the generalized double well
potential.

In Section 2, we give a brief review of our convergent iterative
method for the potential (1.6). To ensure rapid convergence, we
have established a rather effective theorem, called Hierarchy
Theorem, provided a certain inequality can be satisfied. This
inequality is proved in Section 3 for $a=2$ (i.e., the potential
(1.2) in one-dimension), and in the Appendix for an arbitrary
positive $a>a_c\cong 0.664$. Some of the pertinent numerical
results are given in Section 4.

%\newpage

\section*{\bf 2. Trial Function and Iterative Equations}
\setcounter{section}{2} \setcounter{equation}{0}

To construct a good trial function $\phi(x)$ for the groundstate
wave function, we follow the same steps developed for the
double-well potential[2-5]. The main components of $\phi(x)$ are
constructed by using the Schroedinger equation (1.5) and extracting
the first two terms of its perturbation series expansion of
$\psi(x)$ and $E$ in $g^{-1}$.

Write
$$
\psi(x)=e^{-gS(x)}\eqno(2.1)
$$
and set
$$
gS(x)=gS_0(x)+S_1(x)+g^{-1}S_2(x)+\cdots.\eqno(2.2)
$$
Correspondingly,
$$
E=gE_0+E_1+g^{-1}E_2+\cdots.\eqno(2.3)
$$
Substituting (2.1)-(2.3) into (1.5), we find
$$
S_0'=(x^2-1)\sqrt{x^2+a},\eqno(2.4)
$$
and
$$
S_1'S_0'=\frac{1}{2}S''_0-E_0.\eqno(2.5)
$$
Since the left side of (2.5) is zero at $x=1$,
$$
E_0=\frac{1}{2}S_0''(1)=\sqrt{1+a}.\eqno(2.6)
$$
Thus,
$$
S_1'=\frac{x(3x^2+2a-1)-2\sqrt{1+a}\sqrt{x^2+a}}{2(x^2-1)(x^2+a)}.\eqno(2.7)
$$
(Throughout the paper, prime denotes $d/dx$.) Next, introduce
$$
\phi_+(x)\equiv exp[-gS_0(x)-S_1(x)]\eqno(2.8)
$$
and
$$
\phi_-(x)\equiv exp[-gS_0(-x)-S_1(x)].\eqno(2.9)
$$
We note that $\phi_+(x)$ and $\phi_-(x)$ are completely defined,
except for a common arbitrary normalization factor. The trial
function $\phi(x)$ is an even function of $x$, defined by
$$
\phi(x)=\phi(-x)=\left\{
\begin{array}{ll}
\phi_+(x)+\Gamma~\phi_-(x)
&~~~~~~{\sf for}~~0 \leq x<1,\\
\{1+[\Gamma \phi_-(1)/\phi_+(1)]\}\phi_+(x)&~~~~~~{\sf for}~~x>1,
\end{array}
\right.\eqno(2.10)
$$
with
$$
\Gamma=-\frac{\phi_+'(0)}{\phi_-'(0)}=\frac{ga-\sqrt{1+a}}{ga+\sqrt{1+a}}~.\eqno(2.11)
$$
In the following, we assume
$$
g>\frac{1}{a}\sqrt{1+a}~,\eqno(2.12)
$$
and therefore
$$
\Gamma >0.\eqno(2.13)
$$
By construction, $\phi'(0)=0$. The functions $\phi(x)$ and
$\phi'(x)$ are continuous everywhere. Since $\psi(x)$ and $\phi(x)$
are both even in $x$, we need only consider
$$
x\geq 0\eqno(2.14)
$$
in the following.

By differentiation, $\phi_+$ and $\phi$ satisfy two Schroedinger
equations
$$
-\frac{1}{2}\phi_+''+(V+u)\phi_+ =gE_0\phi_+\eqno(2.15)
$$
and
$$
-\frac{1}{2}\phi''+(V+w)\phi =gE_0\phi,\eqno(2.16)
$$
where
$$
u(x)=\frac{1}{2}(S_1'^2-S_1'')\eqno(2.17)
$$
and
$$
w(x)=u(x)+\hat{g}(x)\eqno(2.18)
$$
with
$$
\hat{g}(x) = \left\{
\begin{array}{ll}
gE_0\frac{2\Gamma\phi_-}{(\phi_++\Gamma\phi_-)}
&~~~~~~{\sf for}~~0 \leq x<1\\
0&~~~~~~{\sf for}~~x>1.
\end{array}
\right.\eqno(2.19)
$$
Thus, for
$$
0 \leq x<1,\eqno(2.20)
$$
$$
\hat{g}'(x)=gE_0\frac{2\Gamma\phi_+^2}{(\phi_++\Gamma\phi_-)^2}\bigg(\frac{\phi_-}{\phi_+}\bigg)'.\eqno(2.21)
$$
From (2.8)-(2.9), it follows that
$$
\frac{\phi_-}{\phi_+}=e^{-2gS_0(0)}\cdot e^{2gS_0(x)}.\eqno(2.22)
$$
Within the range (2.20),
$$
\bigg(\frac{\phi_-}{\phi_+}\bigg)'=\bigg(\frac{\phi_-}{\phi_+}\bigg)2g(x^2-1)\sqrt{x^2+a}<0\eqno(2.23)
$$
and therefore
$$
\hat{g}'(x)<0.\eqno(2.24)
$$
Thus, $\hat{g}(x)$ is a positive decreasing function of $x$. At
$x=1$, $\hat{g}$ has a discountinuity, decreasing to zero for $x>1$.

For the one-dimensional case of the generalized sombrero shaped
potential (1.2)-(1.3), we have
$$
N=1,~~~r_0=1~~~{\sf and}~~~a=2.\eqno(2.25)
$$
The corresponding potential is
$$
V(x)=\frac{g^2}{2}(x^2-1)^2(x^2+2).\eqno(2.26)
$$
In the next section, it will be proved that for this potential, at
all finite $x>0$, we have
$$
u(x)>0\eqno(2.27)
$$
and
$$
u'(x)<0.\eqno(2.28)
$$
Thus
$$
w(x)>0\eqno(2.29)
$$
and
$$
w'(x)<0.\eqno(2.30)
$$
Furthermore, $w(\infty)=0$. [The extension to the potential (1.6)
with $a\neq 2$ will be discussed in the Appendix.]

Once the condition $w'(x)<0$ is established, we can apply the
Hierarchy Theorem[3,4], as we shall discuss.

Define
$$
H_0=-\frac{1}{2}\frac{d^2}{dx^2}+V(x)+w(x)\eqno(2.31)
$$
and write (2.16) as
$$
(H_0-gE_0)\phi(x)=0.\eqno(2.32)
$$
The Schroedinger equation we would like to solve is (1.5), which can
be written as
$$
(H_0-gE_0)\psi(x)=(w(x)-{\cal E})\psi(x)\eqno(2.33)
$$
where
$$
{\cal E}=gE_0-E.\eqno(3.34)
$$
Multiplying (2.33) by $\phi$ and (2.32) by $\psi$. Their
difference gives
$$
-\frac{1}{2}(\phi\psi'-\psi\phi')'=(w-{\cal
E})\phi\psi.\eqno(2.35)
$$
Hence, the ratio
$$f=\psi/\phi\eqno(2.36)
$$
satisfies
$$
-\frac{1}{2}(\phi^2f')'=(w-{\cal E})\phi^2f.\eqno(2.37)
$$
Its integral over all $x$ gives
$$
{\cal E}=\int\limits^\infty_0 w \phi^2
f~dx\bigg{/}\int\limits^\infty_0
 \phi^2 f~dx.\eqno(2.38)
 $$
 Eqs. (2.37) and (2.38) can then be solved by considering the
 iterative series $\{f_n(x)\}$ and $\{{\cal E}_n\}$, with
$$
-\frac{1}{2}(\phi^2f_n')'=(w-{\cal E}_n)\phi^2f_{n-1}\eqno(2.39)
$$
and therefore
$$
{\cal E}_n=\int\limits_0^{\infty}
w\phi^2f_{n-1}dx\bigg{/}\int\limits_0^{\infty}
\phi^2f_{n-1}dx.\eqno(2.40)
$$

We differentiate two different sets of boundary conditions:
$$
{\rm (I)}~~~f_n(\infty)=1~~{\sf for~~all}~~n\eqno(2.41)
$$
or
$$
{\rm (II)}~~~f_n(0)=1~~{\sf for~~all}~~n.\eqno(2.42)
$$
Thus, in case {\rm (I)}
$$
f_n(x)=1-2\int\limits_x^{\infty}
\frac{dy}{\phi^2(y)}\int\limits_y^\infty (w(z)-{\cal
E}_n)\phi^2(z)f_{n-1}(z)dz\eqno(2.43)
$$
and correspondingly in case {\rm (II)}
$$
f_n(x)=1-2\int\limits_0^{x} \frac{dy}{\phi^2(y)} \int\limits_0^{y}
(w(z)-{\cal E}_n)\phi^2(z)f_{n-1}(z)dz.\eqno(2.44)
$$
In case {\rm (I)}, it can be readily verified that because
$f_n(\infty)=1$ and
$$
f_n'<0,\eqno(2.45)
$$
we have
$$
f_n(0)>f_n(x)>f_n(\infty)=1.\eqno(2.46{\rm I})
$$
In case {\rm (II)}, we assume $w(x)$ to be not too large so that
(2.44) is consistent with
$$
f_n(x)>0~~{\sf for~~all}~~x
$$
and therefore
$$
f_n(0)=1>f_n(x)>f_n(\infty)>0.\eqno(2.46{\rm II})
$$
As we shall see, these two boundary conditions {\rm (I)} and {\rm
(II)} produce sequences that have very different behavior. Yet,
they also share a number of common properties.

Now, assuming that $u'(x)<0$ (section 3) and hence $w'(x)<0$ (by
(2.18) and (2.28)), we have from refs.[3,4]

{\it Hierarchy Theorem.} {\rm (I)} With the boundary condition
$f_n(\infty)=1$, we have for all $n$
$$
{\cal E}_{n+1} > {\cal E}_n\eqno(2.47)
$$
and
$$
\frac{d}{dx}\left( \frac{f_{n+1}(x)}{f_n(x)}\right)<0 ~~~~~{\sf
at~~any}~~~x>0.\eqno(2.48)
$$
Thus, the sequences $\{{\cal E}_n\}$ and $\{f_n(x)\}$ are all
monotonic, with
$$
{\cal E}_1<{\cal E}_2<{\cal E}_3<\cdots\eqno(2.49)
$$
and
$$
1<f_1(x)<f_2(x)<f_3(x)<\cdots\eqno(2.50)
$$
at all finite $x$.

{\rm (II)} With the boundary condition $f_n(0)=1$, we have for all
odd $n=2l+1$ an ascending sequence
$$
{\cal E}_1<{\cal E}_3<{\cal E}_5<\cdots,\eqno(2.51)
$$
but for all even $n=2m$, a descending sequence
$$
{\cal E}_2>{\cal E}_4>{\cal E}_6>\cdots.\eqno(2.52)
$$
Furthermore, between any even $n=2m$ and any odd $n=2l+1$
$$
{\cal E}_{2m}>{\cal E}_{2l+1}.\eqno(2.53)
$$
Likewise, at any $x$, for any even $n=2m$
$$
\frac{d}{dx}\left(
\frac{f_{2m+1}(x)}{f_{2m}(x)}\right)<0,\eqno(2.54)
$$
whereas for any odd $n=2l+1$
$$
\frac{d}{dx}\left(
\frac{f_{2l+2}(x)}{f_{2l+1}(x)}\right)>0.\eqno(2.55)
$$
The groundstate energy $E$ of the original Hamiltonian
$$
H= -\frac{1}{2}\frac{d^2}{dx^2} +V(x)\eqno(2.56)
$$
is the limit of the sequence $\{E_n\}$ with
$$
E_n=E_0-{\cal E}_n.\eqno(2.57)
$$
Hence the boundary condition $f_n(\infty)=1$ yields a sequence
$$
E_1>E_2>E_3>\cdots >E\eqno(2.58)
$$
with each member $E_n$ an {\it upper} bound of $E$, similar to the
usual variational method. On the other hand, with the boundary
condition $f_n(0)=1$, while the sequence of its odd members
$n=2l+1$ yields a similar one, like (2.58) with
$$
E_1>E_3>E_5>\cdots >E,\eqno(2.59)
$$
its even members $n=2m$ satisfy
$$
E_2<E_4<E_6<\cdots <E.\eqno(2.60)
$$
It is unusual to have an iterative sequence of lower bounds of the
eigenvalue $E$. In both cases we have
$$
\lim\limits_{n\rightarrow\infty} E_n=E.\eqno(2.61)
$$

\newpage

\section*{\bf 3. Proof of $u'(x)<0$}
\setcounter{section}{3} \setcounter{equation}{0}

In this section we shall establish  $u'(x)<0$ for the case $a=2$
[i.e., the one-dimensional case of the generalized sombrero
potential (1.2)]. The general case of arbitrary positive $a$ will be
discussed in the Appendix.

Write the potential $V(x)$ of (1.6) as
$$
V(x)=g^2v(x).\eqno(3.1).
$$
When $a=2$,
$$
v(x)=\frac{1}{2}(x^2-1)^2(x^2+2).\eqno(3.2)
$$
The functions $S_0(x)$ and $S_1(x)$ introduced in Section 2 are
$$
S_0(x)=\frac{x}{4}(x^2-1)\sqrt{x^2+2}-\frac{3}{2}\ln
(x+\sqrt{x^2+2})\eqno(3.3)
$$
and
$$
S_1(x)=\ln (x+1)+\frac{1}{4}\ln (x^2+2)+\frac{1}{2}\ln
\frac{2+x+\sqrt{3(x^2+2)}}{2-x+\sqrt{3(x^2+2)}}.\eqno(3.4)
$$
In accordance with (2.17), we find
$$
u(x)=\frac{3}{8}~\frac{25x^8+150x^6+393x^4+408x^2+144}{(x^2+2)^2(A+B)}\eqno(3.5)
$$
with
$$
A=5x^6+10x^4+21x^2+12\eqno(3.6)
$$
and
$$
B=8\sqrt{3}x(x^2+1)\sqrt{x^2+2}.\eqno(3.7)
$$
Its derivative is
$$
u'(x)=-\frac{3}{8}~\frac{C_1\sqrt{x^2+2}+C_2}{\sqrt{x^2+2}(\alpha+\beta\sqrt{x^2+2})^2}\eqno(3.8)
$$
with
\newpage
$$
C_1=250x^{17}+3000x^{15}+16740x^{13}+49880x^{11}+83706x^9+77952x^7
$$
$$
+34008x^5+2880x^3-1152x,\eqno(3.9)
$$

$$
C_2=8\sqrt{3}(322x^{12}+2236x^{10}+6322x^8+9672x^6+8904x^4
$$
$$
+4800x^2+1152),\eqno(3.10)
$$

$$
\alpha=(x^2+2)^2A\eqno(3.11)
$$
and
$$
\beta=(x^2+2)^{\frac{3}{2}}B=8\sqrt{3}x(x^2+1)(x^2+2)^2.\eqno(3.12)
$$
Because at all $x$
$$
(4800x^2+1152)8\sqrt{3}>1152x\sqrt{x^2+2},\eqno(3.13)
$$
we have
$$
C_1\sqrt{x^2+2}+C_2>0\eqno(3.14)
$$
and therefore
$$
u'<0.\eqno(3.15)
$$
From (3.5), we see that $u>0$ and therefore Hierarchy Theorem is
applicable. The general case when $V(x)$ is given by (1.6) will be
discussed in the Appendix. As we shall see, for $V(x)=g^2v(x)$,
$$
v(x)=\frac{1}{2}(x^2-1)^2(x^2+a)\eqno(3.16)
$$
and
$$
a>a_c\cong 0.664,\eqno(3.17)
$$
the corresponding $u(x)$ also satisfies (3.15); i.e. $u'(x)<0$ for
all $x>0$, and that ensures the applicability of the Hierarchy
Theorem.

\section*{\bf 4. Numerical Results and Discussions}
\setcounter{section}{4} \setcounter{equation}{0}

Throughout this section we denote the $n^{th}$ order iterative
solution as
$$
\psi_n(x)=\psi_0(x)f_n(x)\eqno(4.1)
$$
with $f_n(x)$ the solution of (2.39) and $\psi_0(x)$ related to
$\phi(x)$ of (2.10) by
$$
\psi_0(x)=\phi(x)/\phi(0)\eqno(4.2)
$$
so that
$$
\psi_0(0)=1.\eqno(4.3)
$$
\\

\noindent {\bf 4.1 The case $a=2$ and $g=1$.}

We first discuss the case of $a=2$ and $g=1$. The resultant series
of energies are shown in Table~1. The energy series for two
different boundary conditions converge in different ways.

{\rm I} . For the boundary condition $f_n(\infty)=1$ we find that
the iterative energy sequence
$$
E_0>E_1>E_2>E_3>\cdots>E\eqno(4.4)
$$
is given numerically by
$$
1.7321>1.0163 > 1.0031 > 1.0005 > 1.0001
>1.0000>\cdots>E=1.\eqno(4.5)
$$

{\bf\rm II}. For the boundary condition $f_n(0)=1$, $E_0$ is still
$1.7321$, but the iterative energy sequence becomes
$$
E_1>E_3>\cdots>E>\cdots>E_4>E_2,\eqno(4.6)
$$
and is given numerically by
$$
1.0163>1.0002>\cdots>E>\cdots>1.0000>0.9981~.\eqno(4.7)
$$
To 4 decimal places $E_4$ is indistinguishable from the exact $E=1$.

In both cases ${\rm I}$ and ${\rm II}$, the zeroth order trial
function $\psi_0(x)$ has a maximum near $x=1$. However, to the
accuracy of Figure 1, for $n=2$ $\psi_2(x)$ is essentially the
same as the exact groundstate wave function $\psi(x)=e^{-x^4/4}$.
The resultant series of wave functions and energies convergence
rapidly to the exact expressions. It is interesting to notice that
although the trial function has its two maxima near $x=\pm 1$, the
iterative series gives the final wave function in the shape of the
exact groundstate wave function with a single maximum at $x=0$.
The rapid change of the shape of the wave function from the trial
one to the exact
solution shows how the iteration procedure works.\\

\noindent {\bf 4.2 The case $g=1$.}

Next we discuss the case of $g=1$ for different values of $a$. The
results of the energy series for the boundary condition ${\rm II}$
are given in Table~2.

As proved in Appendix, to ensure the convergence of the iterative
series the parameter $a$ should satisfy $a>a_c\cong 0.664$. We
have chosen $g=1$, $a=3,~2$ and $1.8$. The obtained wave functions
are shown respectively in Figures 2, 1 and 3. The behavior of the
final function for $a<2$ is similar to the one at $a=2$, namely
the resultant wave function has only one maximum at $x=0$, while
the trial function has two maxima near $x=\pm 1$. When $a=3$ the
iterative wave function retains a similar shape as the trial
function with its maxima at $x$ near $\pm 1$, indicating that the
exact groundstate wave function also has its maximum at $x$ near
$\pm 1$, like the trial function
when $g=1$ and $a=3$.\\

\noindent {\bf 4.3 The case $a=2$.}

Next we discuss the case $a=2$ for different values of $g$. The
results of the energy series for the boundary condition ${\rm II}$
are given in Table~3.

According to (2.12), for $a=2$ the allowed values of $g$ is
$g>0.866$. The wave functions for $a=2$, $g=3,~1$ and $0.88$ are
shown respectively in Figures 4, 1 and 5. The behavior of the
final function for $g<1$ is similar to the one at $g=1$, namely
the resultant wave function has only one maximum at $x=0$, while
the trial function has two maxima near $x=\pm 1$. When $g=3$ the
iterative wave function retains a similar shape as the trial
function with its maxima at $x$ near $\pm 1$, indicating that the
exact groundstate wave function also has its maximum at $x$ near
$\pm 1$, like the trial function
when $g=3$ and $a=2$.\\

\newpage

{\bf Table 1. Eigenvalues of groundstates for $g=1$ and $a=2$} \\

\begin{tabular}{|c|c|c|c|c|c|c|c|}
  \hline
 & $E_0$ & $E_1$ & $E_2$ & $E_3$ & $E_4$ &$E_5$\\
  \hline
${\rm I}$ & 1.7321 & 1.0163 & 1.0031 & 1.0005 & 1.0001 & 1.0000  \\
\hline
${\rm II}$ & 1.7321 & 1.0163 & 0.9981 & 1.0002 & 1.0000 & 1.0000 \\
\hline
\end{tabular}

\vspace{1cm}

{\bf Table 2. Eigenvalues of groundstates for $g=1$,} \\
\hspace*{1.5cm}using ${\rm II}$ with $f_n(0)=1$ as the boundary condition \\

\begin{tabular}{|c|c|c|c|c|c|c|c|}
  \hline
$a$ & $E_0$ & $E_1$ & $E_2$ & $E_3$ & $E_4$ &$E_5$\\
  \hline
1.8 & 1.6733 & 0.9558 & 0.9418 & 0.9432 & 0.9431 & 0.9431 \\
\hline
2 & 1.7321 & 1.0163 & 0.9981 & 1.0002 & 1.0000 & 1.0000 \\
\hline
3 & 2.0000 & 1.2974 & 1.2602 & 1.2659 & 1.2651 & 1.2652 \\
\hline
\end{tabular}

\vspace{1cm}

{\bf Table 3. Eigenvalues of groundstates for $a=2$} \\
\hspace*{1.5cm}using ${\rm II}$ with $f_n(0)=1$ as the boundary condition \\

\begin{tabular}{|c|c|c|c|c|c|c|c|}
  \hline
$g$ & $E_0$ & $E_1$ & $E_2$ & $E_3$ & $E_4$ &$E_5$\\
\hline
 0.88 & 1.5242 & 0.8633 & 0.8517 & 0.8528 & 0.8527 & 0.8527\\
\hline
1 & 1.7321 & 1.0163 & 0.9981 & 1.0002 & 1.0000 & 1.0000 \\
\hline
2 & 3.4641 & 2.6934 & 2.6375 & 2.6465 & 2.6455 & 2.6456 \\
\hline
3 & 5.1962 & 4.5786 & 4.5562 & 4.5591 & 4.5589 & 4.5589 \\
\hline
\end{tabular}

\section*{\bf References}

\noindent [1] R. Jackiw, Private communications.

\noindent [2] R. Friedberg, T. D. Lee, W. Q. Zhao and A. Cimenser

~~~~~~~~~~~Ann. Phys. 294 (2001), 67\\
\noindent [3] R. Friedberg and T. D. Lee, Ann. Phys. 308 (2003),
263\\
\noindent [4] R. Friedberg and T. D. Lee, Ann. Phys. 316(2005), 44\\
\noindent [5] R. Friedberg, T. D. Lee and W. Q. Zhao, Ann. Phys. 321 (2006), 1981\\
\noindent [6] G. 't Hooft and S. Nobbenhuis, Class. Quant. Grav. 23(2006), 3819\\

\newpage

\section*{\bf Appendix}
\setcounter{section}{5} \setcounter{equation}{0}

In this Appendix we give the proof of the convergence of the
iterative series for the generalized double-well potential, eq.(1.6)
$$
V(x) = g^2 v(x)~~~{\sf
and}~~~v(x)=\frac{1}{2}(x^2-1)^2(x^2+a),\eqno(A.1)
$$
where $a$ is an arbitrary constant.

Following the steps described in Section 2, from (2.4), (2.7) and
(2.17) we have
$$
u(x)=\frac{\alpha-8\sqrt{x^2+a}~\beta}{8(x^2-1)^2(x^2+a)^2}\eqno(A.2)
$$
where
$$
\alpha=15x^6+6(3a-1)x^4+(8a^2+12a+7)x^2+8a^2+2a\eqno(A.3)
$$
and
$$
\beta=\sqrt{1+a}~x(3x^2+2a-1).\eqno(A.4)
$$
Integrating (2.4) and (2.7) gives
$$
S_0(x)=\frac{1}{4}x(\sqrt{x^2+a})^3-\bigg(\frac{a}{8}+\frac{1}{2}\bigg)x\sqrt{x^2+a}
$$
$$
-a\bigg(\frac{a}{8}+\frac{1}{2}\bigg)\ln
(x+\sqrt{x^2+a})\eqno(A.5)
$$
and
$$
S_1(x)=\ln (x+1)(x^2+a)^{1/4}+\frac{1}{2}\ln
\frac{\sqrt{a+1}\sqrt{x^2+a}+a+x}{\sqrt{a+1}\sqrt{x^2+a}+a-x}.\eqno(A.6)
$$
The corresponding trial function (2.10) satisfies the Schroedinger
equation (2.16) with $w(x)$ expressed by (2.18) and (2.19).

According to the Hierarchy Theorem proved in Ref.[3,4] the
iterative series is convergent if the potential $w(x)$, defined by
(2.18), satisfies the conditions $w(x)>0$ and $w'(x)<0$. As we
shall prove, when
$$
a>a_c \cong 0.664,\eqno(A.7)
$$
the potential $u(x)$, defined by (2.17), satisfies $u(x)>0$ and
$u'(x)<0$. Furthermore, when $g$ satisfies (2.12), i.e.
$$
a>a_g=\frac{1+\sqrt{1+4g^2}}{2g^2}\eqno(A.8)
$$
we have $\Gamma>0$ from (2.13); therefore $\hat{g}(x)>0$ and
$\hat{g}'(x)<0$. Combining these results we have
$$
w(x)>0~{\sf and}~w'(x)<0~~~~~~~{\sf
when}~a>a_m=\max(a_c,~a_g).\eqno(A.9)
$$
This condition ensures the convergent iterative solution for the
groundstate of the generalized double-well potential (A.1) (i.e.
(1.6)), in accordance with the Hierarchy theorem.\\

\noindent {\bf  Range of parameter $a$}

As mentioned above, to obtain a convergent iterative series the
potential $u(x)$ in (A.2) should satisfy
$$
u(x)>0~~~{\sf and}~~~u'(x)<0.\eqno(A.10)
$$
For $a>0$, introducing
$$
\gamma_{\pm} \equiv \alpha \pm 8\sqrt{x^2+a}~\beta\eqno(A.11)
$$
with $\alpha$ and $\beta$ given by (A.3)-(A.4), we have
$$
\gamma_+\gamma_-\equiv
(x^2-1)^2\gamma=\alpha^2-64(x^2+a)~\beta^2.\eqno(A.12)
$$
Thus, $u(x)$ can be expressed as
$$
u=\frac{\gamma_-}{8(x^2+a)^2(x^2-1)^2}\eqno(A.13)
$$
and
$$
u=\frac{1}{8(x^2+a)^2}
\bigg(\frac{\gamma}{\gamma_+}\bigg).\eqno(A.14)
$$
Write
$$
\gamma=\gamma(a,~x)=g_1(x)(15x^4+36ax^2)+g_2\eqno(A.15)
$$
where
$$
g_1(x)=15x^4+18x^2-1\eqno(A.16)
$$
and
$$
g_2=g_2(a,~x^2)=4(141x^4+90x^2+1)a^2+32(9x^2+1)a^3+64a^4>0.\eqno(A.17)
$$

Consider only the first quadrant in $(x,~a)$-plane with $x,~a$
positive. Define $x_0$ to be the point satisfying $g_1(x_0)=0$. For
$x>x_0$, $g_1(x)>0$ and therefore, $\gamma>0$. For $x<x_0$,
$g_1(x)<0$ and
$$
\frac{\partial}{\partial a}\bigg(\frac{\gamma}{a^2}\bigg)>0~~{\sf
for}~~a>0.\eqno(A.18)
$$
Since $(\gamma/a^2) \rightarrow -\infty$ when $a\rightarrow 0$ and
$(\gamma/a^2) \rightarrow \infty$ when $a \rightarrow \infty$, at a
fixed $x<x_0$ by varying the parameter $a$ from $0$ to $\infty$,
$\gamma(a,~x)=0$ only once. We define a curve $(G)$ in $x$,
satisfying $\gamma(a,~x)=0$ and a curve $(B)$ satisfying
$\beta(a,~x)=0$ as shown in Fig.~A1. It can be seen that above the
curves (B) and $(G)$, $\beta$ and $\gamma$ are both positive. The
curve (B) is above the curve (G). Therefore above the curves (B)
$\gamma_+$, $\gamma$ and $u$ are all positive. Below the curve (B)
$\beta$ is negative; therefore $\gamma_-$ and $u$ are positive. Thus
we can conclude that $u(x)>0$ everywhere for $a>0$.

By differentiating $u$, we have
$$
\bigg(\frac{\partial u}{\partial x}\bigg)_a=u'(x)\equiv
\frac{\tilde{\gamma}_-}{8(x^2+a)^3(x^2-1)^3},\eqno(A.19)
$$
where
$$
\tilde{\gamma}_-=\tilde{\alpha}-8\sqrt{x^2+a}~\tilde{\beta}\eqno(A.20)
$$
with
$$
\tilde{\alpha}=[x^4+(a-1)x^2-a]\alpha'-[8x^3+4(a-1)x]\alpha
$$
$$
=(-30x^9-6x^7-42x^5+14x^3)+(-42x^7-162x^5+18x^3-6x)a
$$
$$
+(-48x^5-144x^3)a^2+(-16x^3-48x)a^3\eqno(A.21)
$$
and
$$
\tilde{\beta}=(x^2+a)(x^2-1)\beta'-[7x^3+(4a-3)x]\beta
$$
$$
=\sqrt{a+1}\bigg[(-12x^6+6x^4-2x^2)+(-15x^4-2x^2+1)a+(-6x^2-2)a^2\bigg].\eqno(A.22)
$$
Defining
$$
\tilde{\gamma}_{\pm} \equiv \tilde{\alpha} \pm
8\sqrt{x^2+a}~\tilde{\beta}\eqno(A.23)
$$
we have
$$
\tilde{\gamma}_+\tilde{\gamma}_-=(x^2-1)^3\tilde{\gamma}
$$
$$
=\tilde{\alpha}^2-64(x^2+a)~\tilde{\beta}^2.\eqno(A.24)
$$
Therefore
$$
u'=\frac{1}{8(x^2+a)^3}\bigg(\frac{\tilde{\gamma}}{\tilde{\gamma}_+}\bigg).\eqno(A.25)
$$
It is straightforward to show after some derivation that
$\tilde{\gamma}$ is a polynomial of $x^2$ and $a$. Defining
$$
\tilde{\gamma}\equiv \sum\limits_{\lambda=0}^6
\tilde{\Gamma}_\lambda x^{2\lambda}\eqno(A.26)
$$
where $\tilde{\Gamma}_\lambda$ are polynomials of $a$:
\begin{eqnarray*}
~~~~~~~~~~~~~~~~~~\tilde{\Gamma}_0&=&a^3[64-192a+256a^3]\\
\tilde{\Gamma}_1&=&a^2[-228-1152a^2+1536a^3]\\
\tilde{\Gamma}_2&=&a[168+1068a-960a^2+3648a^3]\\
\tilde{\Gamma}_3&=&60-504a+4500a^2+4992a^3\\
\tilde{\Gamma}_4&=&-180+8568a+4644a^2\\
\tilde{\Gamma}_5&=&3060+2520a\\
\tilde{\Gamma}_6&=&900.~~~~~~~~~~~~~~~~~~~~~~~~~~~~~~~~~~~~~~~~~~~~~~~~~~~~~~~(A.27)
\end{eqnarray*}
For small $a$, it is convenient to introduce
$$
z\equiv x^2/a,~~~{\sf i.e.}~~~x^2=az.\eqno(A.28)
$$
We can plot three curves on the $(z,~a)$-plane : the curve
$(\tilde{A})$: $\tilde{\alpha}=0$, $(\tilde{B})$: $\tilde{\beta}=0$
and $(\tilde{G})$: $\tilde{\gamma}=0$. The regions above the three
curves correspond to $\tilde{\alpha}<0$ $\tilde{\beta}<0$ and
$\tilde{\gamma}>0$, respectively (see Fig.~A2 for details). Above
the critical point $C$, i.e. $a>a_c\cong 0.664$ we have $u'<0$ from
(A.23) and (A.25). Therefore this is the condition for $u>0$ and
$u'<0$.

\newpage

{\bf Figure Caption}\\

\noindent Fig. 1~  Trial Function $\psi_0(x)$ and Groundstate Wave
 Function $\psi(x)$

~~~~~ for $g=1$ and $a=2$.\\

 \noindent Fig. 2~  Trial Function $\psi_0(x)$ and Groundstate Wave
 Function $\psi(x)$

 ~~~~~ for $g=1$ and $a=3$.\\

 \noindent Fig. 3~  Trial Function $\psi_0(x)$ and Groundstate Wave
 Function $\psi(x)$

 ~~~~~ for $g=1$ and $a=1.8$.\\

\noindent Fig. 4~  Trial Function $\psi_0(x)$ and Groundstate Wave
 Function $\psi(x)$

 ~~~~~ for $g=3$ and $a=2$.\\

 \noindent Fig. 5~  Trial Function $\psi_0(x)$ and Groundstate Wave
 Function $\psi(x)$

 ~~~~~ for $g=0.88$ and $a=2$.\\

\noindent  Fig. A1~  Curves for $\beta=0$ and
 $\gamma=0$.\\

\noindent Fig. A2~  Curves for
 $\tilde{\alpha}=0$, $\tilde{\beta}=0$ and $\tilde{\gamma}=0$.\\

%\newpage

\begin{figure}[h]
 \centerline{
\epsfig{file=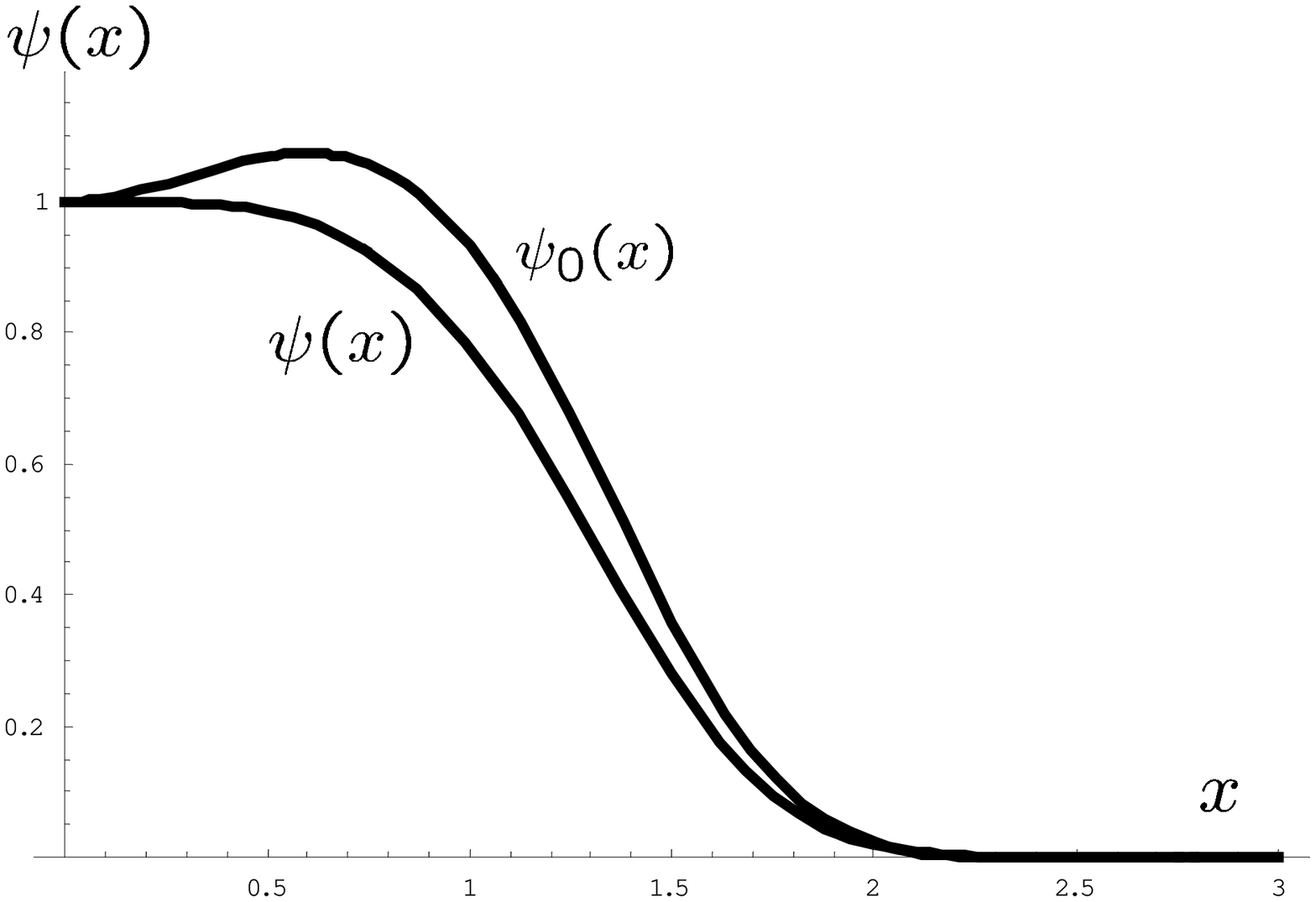, width=7cm, height=11cm, angle=-90}}
\vspace{.5cm}
 \centerline{{\normalsize \sf Fig. 1~  Trial Function $\psi_0(x)$ and Groundstate Wave
 Function $\psi(x)$ }}
  \centerline{{\normalsize \sf for $g=1$ and $a=2$.}}
\end{figure}

\begin{figure}[h]
 \centerline{
\epsfig{file=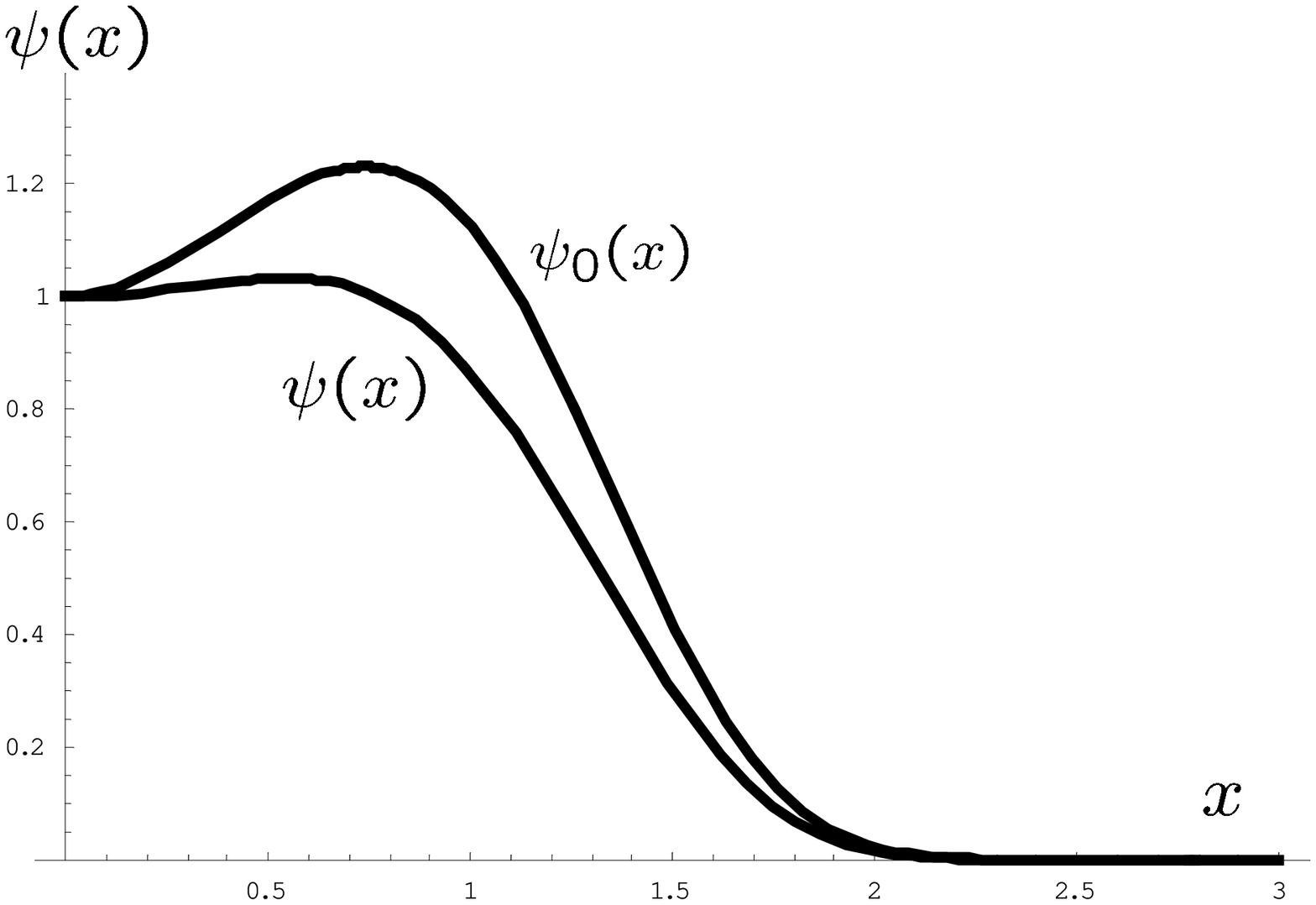, width=7cm, height=11cm, angle=-90}}
\vspace{.5cm}
 \centerline{{\normalsize \sf Fig. 2~  Trial Function $\psi_0(x)$ and Groundstate Wave
 Function $\psi(x)$ }}
  \centerline{{\normalsize \sf for $g=1$ and $a=3$.}}
\end{figure}

%\newpage

\begin{figure}[h]
 \centerline{
\epsfig{file=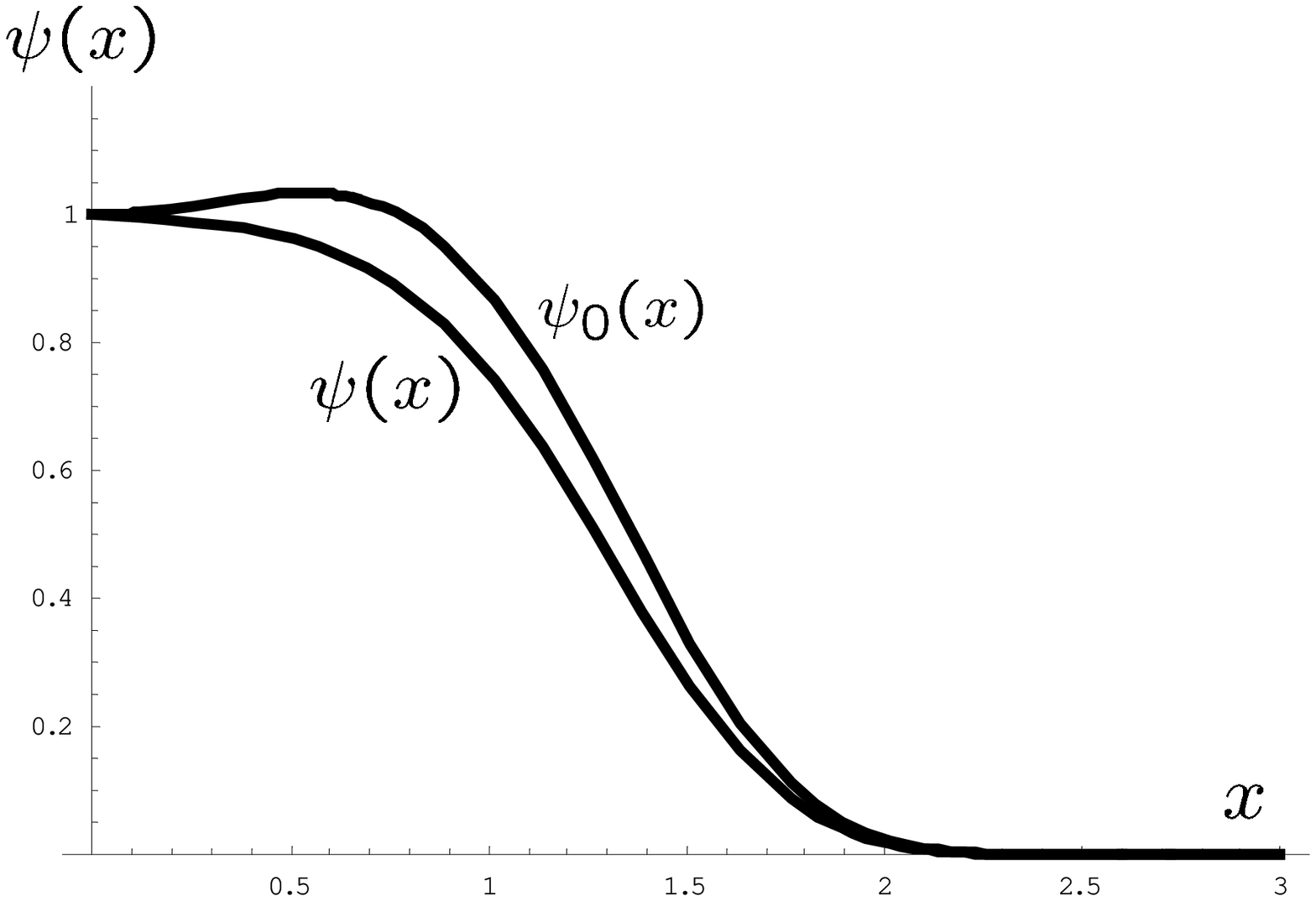, width=7cm, height=11cm, angle=-90}}
\vspace{.5cm}
 \centerline{{\normalsize \sf Fig. 3~  Trial Function $\psi_0(x)$ and Groundstate Wave
 Function $\psi(x)$ }}
  \centerline{{\normalsize \sf for $g=1$ and $a=1.8$.}}
\end{figure}

\begin{figure}[h]
 \centerline{
\epsfig{file=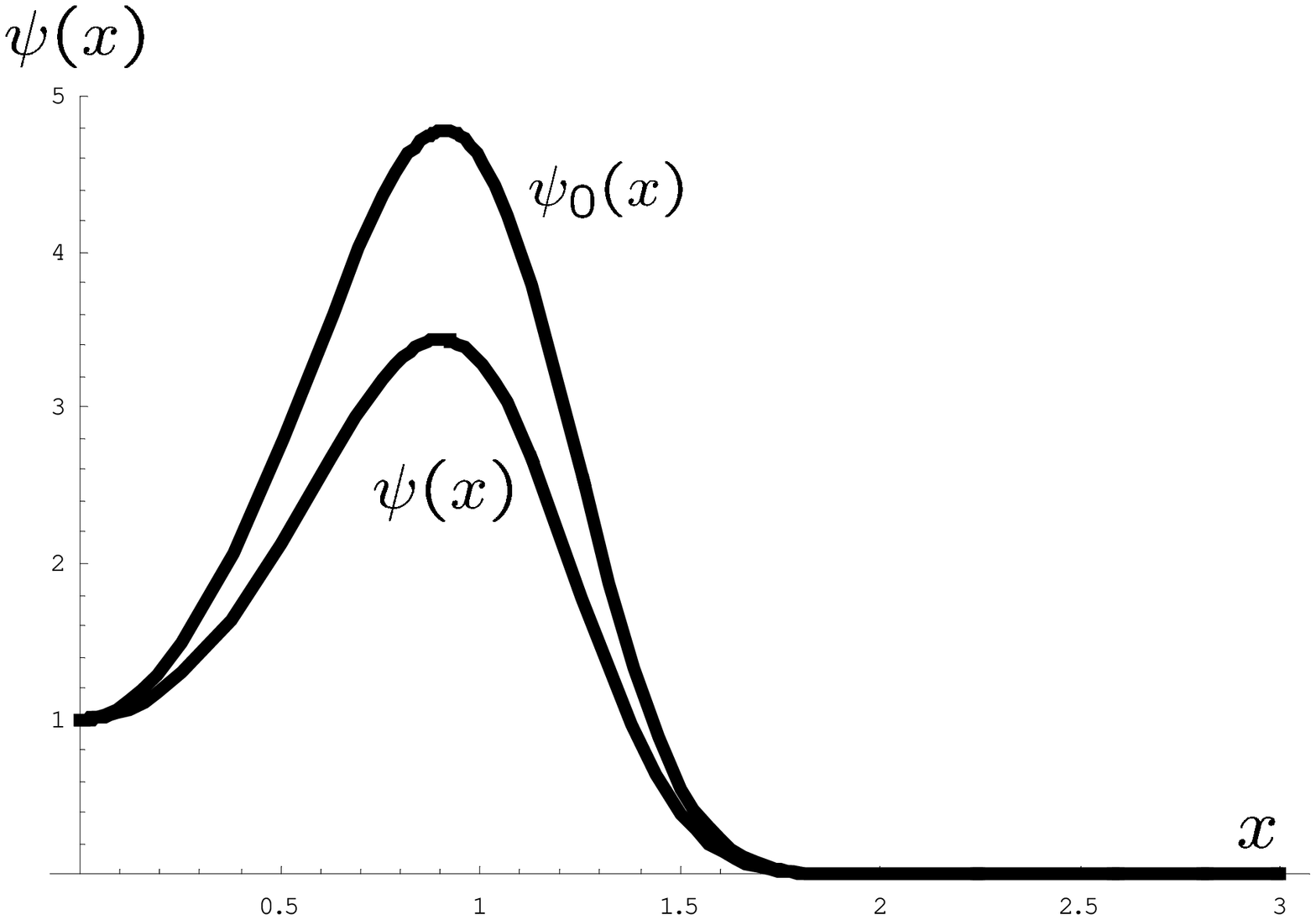, width=7cm, height=11cm, angle=-90}}
\vspace{.5cm}
 \centerline{{\normalsize \sf Fig. 4~  Trial Function $\psi_0(x)$ and Groundstate Wave
 Function $\psi(x)$ }}
  \centerline{{\normalsize \sf for $g=3$ and $a=2$.}}
\end{figure}

\begin{figure}[h]
 \centerline{
\epsfig{file=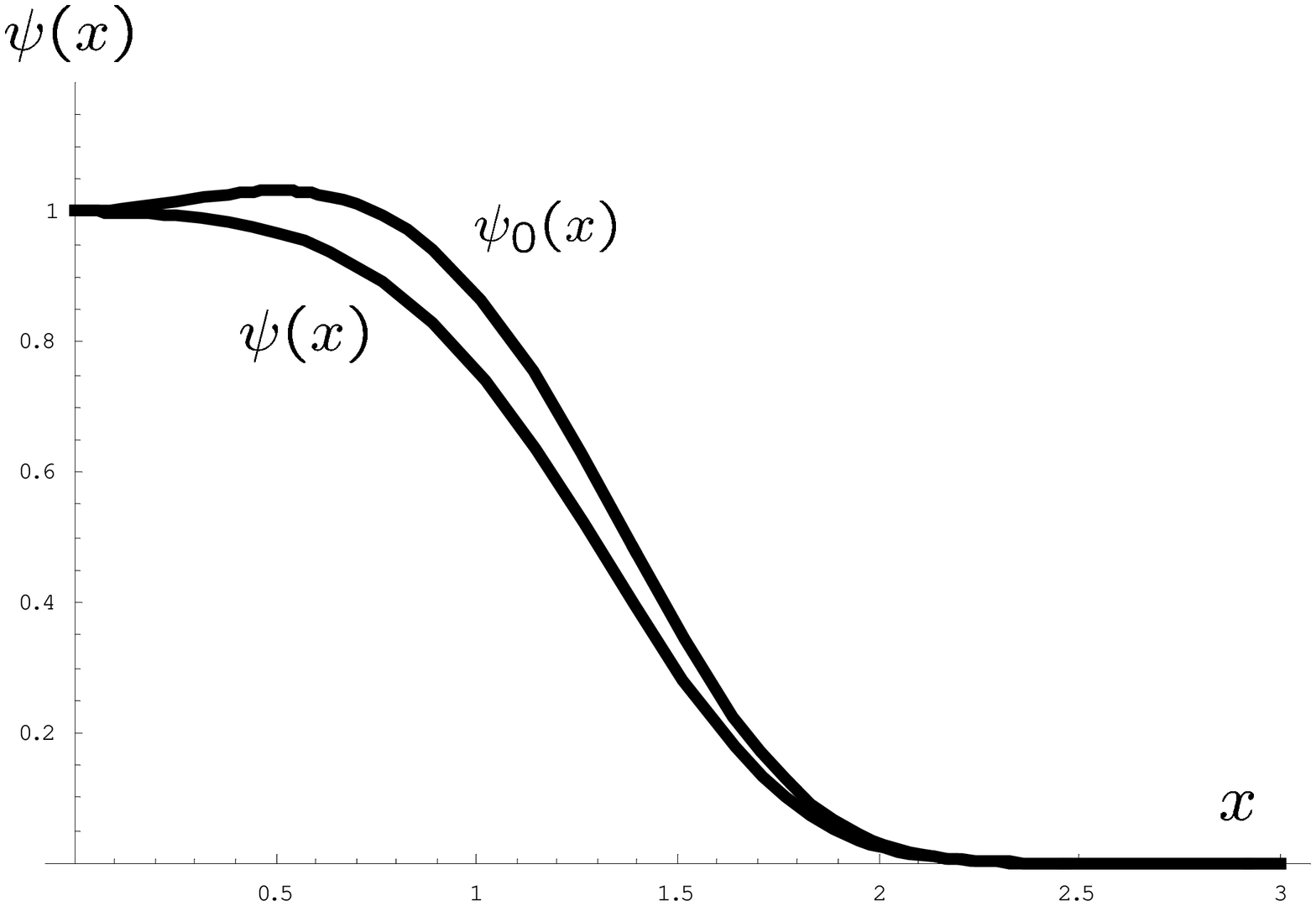, width=7cm, height=11cm, angle=-90}}
\vspace{.5cm}
 \centerline{{\normalsize \sf Fig. 5~  Trial Function $\psi_0(x)$ and Groundstate Wave
 Function $\psi(x)$ }}
  \centerline{{\normalsize \sf for $g=0.88$ and $a=2$.}}
\end{figure}

\begin{figure}[h]
 \centerline{
\epsfig{file=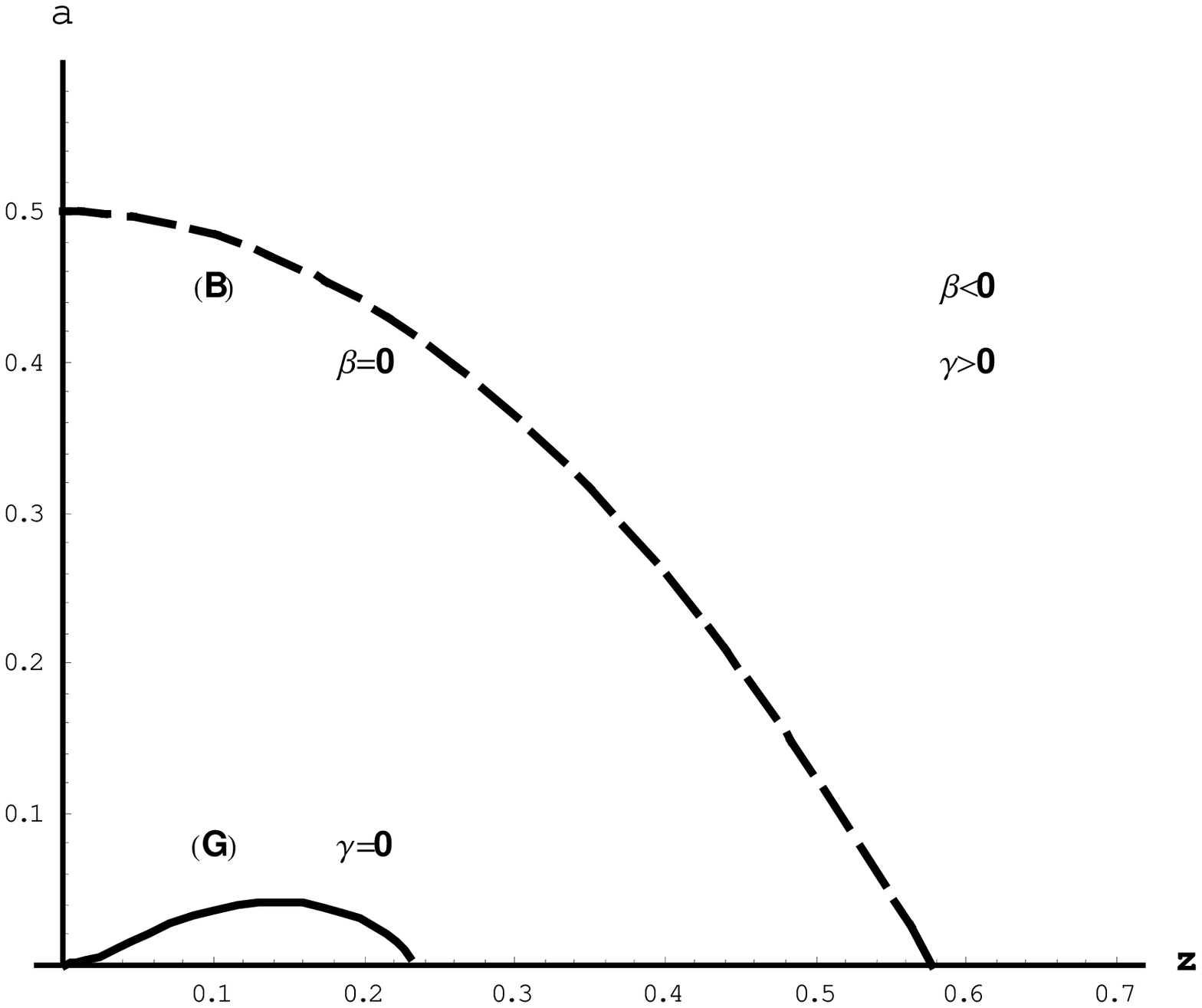, width=9cm, height=11cm, angle=-90}}
\vspace{.5cm}
 \centerline{{\normalsize \sf Fig. A1~  Curves for $\beta=0$ and
 $\gamma=0$..}}
\end{figure}

%\newpage

\begin{figure}[h]
 \centerline{
\epsfig{file=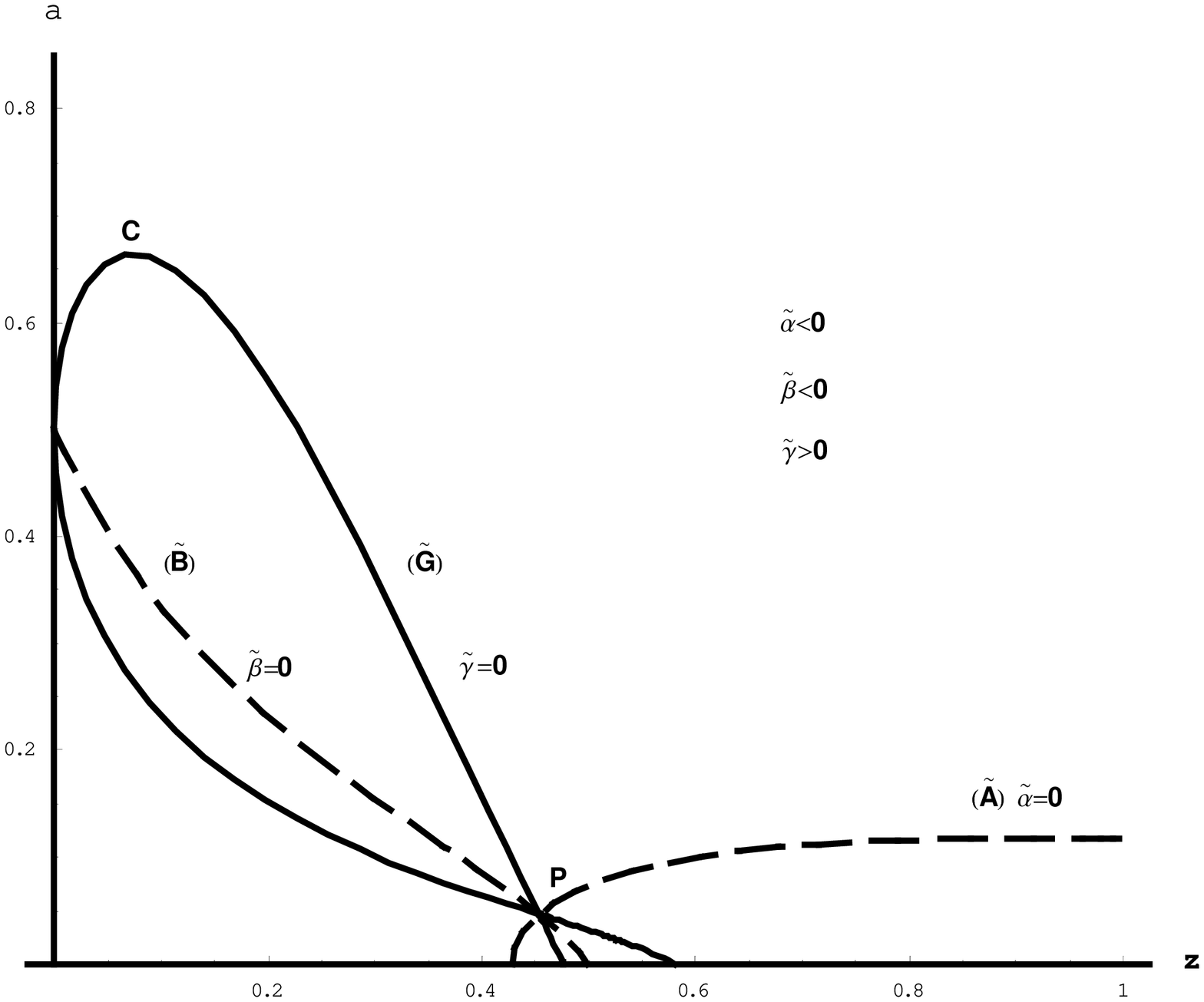, width=9cm, height=11cm, angle=-90}}
\vspace{.5cm}
 \centerline{{\normalsize \sf Fig. A2~  Curves for
 $\tilde{\alpha}=0$, $\tilde{\beta}=0$ and $\tilde{\gamma}=0$.}}
\end{figure}

\end{document}